 \documentclass[aps,prl,twocolumn,superscriptaddress,showpacs,amsmath,amssymb]{revtex4}
\usepackage{graphicx}% Include figure files
\usepackage{dcolumn} % Align table columns on decimal point
\usepackage{bm}      % bold math
\usepackage{verbatim}%

\begin{document}

%\preprint{}
%\preprint{APS/123-QED}

%------------------------- Title --------------------------%
\title
{Free Energy Approach to the Formation of an Icosahedral Structure during the Freezing of Gold Nanoclusters}

%----------------- Author and Affiliation -----------------%

% The \affiliation command should follow the other information
%\author{}
%\thanks{}
%\affiliation{}

\author{ H.-S. Nam }
\email
{hnam@princeton.edu}

\altaffiliation
{Present address: Princeton University, Princeton, New Jersey 08544, USA}
\affiliation
{
Center for Microstructure Science of Materials, School of Materials Science and Engineering,
Seoul National University, Seoul 151-742, South Korea
}

\author{ Nong M. Hwang }

\affiliation
{
School of Materials Science and Engineering, Seoul National University, Seoul 151-742, Korea
}

\author{ B.D. Yu }

\affiliation
{
Department of Physics, University of Seoul, Seoul 130-743, South Korea
}

\author{ D.-Y. Kim }

\affiliation
{
Center for Microstructure Science of Materials, School of Materials Science and Engineering,
Seoul National University, Seoul 151-742, South Korea
}

\author{ J.-K. Yoon }

\affiliation
{
School of Materials Science and Engineering, Seoul National University, Seoul 151-742, Korea
}


%-------------------------- Date --------------------------%
\date{\today}% It is always \today, today,

%------------------------ Abstract ------------------------%
\begin{abstract}

The freezing of metal nanoclusters such as gold, silver, and copper exhibits a novel structural evolution. The formation of the icosahedral (Ih) structure is dominant despite its energetic metastability. The dynamical aspects of the structural transformations, which are eventually responsible for the kinetics, are studied by calculating free energies of gold nanoclusters. The transition barriers have been determined by using the umbrella sampling technique. Our calculations show that the formation of Ih gold nanoclusters is attributed to the lower free energy barrier from the liquid to the Ih phase compared to the barrier from the liquid to the face-centered-cubic crystal phase.

\end{abstract}

%---------------------- PACS number -----------------------%
% 61.46.+w Nanoscale materials: clusters, nanoparticles, nanotubes, and nanocrystals
%          (see also 36.40 Atomic and molecular clusters;
%           for fabrication and characterization of nanoscale materials,
%           see 81.07.-b in materials science)
%
% 36.40.-c Atomic and molecular clusters (see also 61.46.+w in condensed matter)
% 36.40.Ei Phase transitions in clusters
% 36.40.Mr Spectroscopy and geometrical structure of clusters
% 36.40.Qv Stability and fragmentation of clusters
%
% 31.15.-p Calculations and mathematical techniques in atomic and molecular physics
%          (excluding electron correlation calculations)
%          (see also 02.70.-c computational techniques, in mathematical methods in physics)
% 31.15.Qg Molecular dynamics and other numerical methods
%
% 64.70.-p Specific phase transitions
% 64.70.Dv Solid.liquid transitions
% 64.70.Kb Solid.solid transitions
% 64.70.Nd Structural transitions in nanoscale materials
%


%----------------------- Key Words ------------------------%
%\keywords{}

\maketitle

%%%%%%%%%%%%%%%%%%%%%%%%%%%%%%% Body of Paper %%%%%%%%%%%%%%%%%%%%%%%%%%%%%%%%%%

Recently, nanosized metal clusters have been extensively studied as a fundamental element for technological applications, such as nanocatalysts and nanoelectronic devices \cite{MoriartyBinns}. Unlike bulk, at surfaces or inside of metal nanoclusters, atomic bonds may be cut and new bonds formed due to the presence of a nano-size surface or quantum effects. As a result, metal nanoclusters exhibit unique chemical and physical properties distinct from bulk materials. For the controlled growth of low dimensional nanostructures, it is very important to understand the properties of metal nanoclusters, their formation from the liquid state or gas phase, and their chemistry.

In particular, phase transitions of metal nanoclusters have attracted great interest because of novel physical behavior, such as size-dependent melting point depression \cite{Pawlow, Borel}, quasimelting \cite{Iijima, AjayanKrakow}, and dynamic phase transitions \cite{Kunz} in nano-size regions. For instance, small metal clusters have lower melting points compared to the bulk melting point \cite{Pawlow, Borel}. Furthermore, the structure of metal nanoclusters may fluctuate below the melting point under external perturbations \cite{Iijima, AjayanKrakow}. The experimental evidence for the quasimelting of gold nanoclusters was reported by Iijima and Ichihashi \cite{Iijima} through real-time microscopic studies: the gold clusters change from a single crystalline form to a twinned crystalline form based on icosahedral (Ih) or decahedral (Dh) structures, and vice versa, when irradiated with intense electron beams.

Another interesting aspect of phase transitions of nanoclusters is the novel structural evolution of clusters produced from the liquid state or gas phase. Because of a large surface-to-volume ratio in nano-size regions, metal clusters exhibit various structural modifications. For example, for face-centered-cubic (fcc) noble metals such as gold, silver, and copper, nanoclusters of Ih or Dh structures with a fivefold symmetry of noncrystallographic atomic arrangements were dominantly formed as observed in high resolution electron microscopy experiments \cite{MarksMartin, Reinhard, Koga}. The experimental observations indicate that the Ih or Dh structure should be a lower energy state than an fcc structure, by thermodynamic principles, in which the clusters are assumed to adopt the energetically stable thermal-equilibrium structure. In contrast to the experimental observations, previous theoretical calculations suggest that the minimum-energy structure for Au is a truncated octahedron (TOh) rather than the Ih or Dh structure for several hundred atom clusters and the Ih structure is energetically metastable even for small clusters (less than 100 atoms) \cite{ClevelandMS, Baletto}. This novel and important structural transition behavior during the formation of metal nanoclusters thus remains ambiguous.

In order to understand the dynamical aspects of the structural transformations such as the freezing and melting of metal nanoclusters, we have performed free energy calculations of gold nanoclusters using the umbrella sampling technique \cite{Torrie, UnderstandingMD} combined with molecular dynamics simulations. The variation of free energies for the structural transformation was obtained by introducing a crystalline order parameter \cite{Steinhardt}. Our free energy calculations showed that the formation of the Ih structure is attributed to the lower free energy barrier from the liquid to the Ih phase compared to the barrier from the liquid to the fcc crystal phase. This explains why the Ih cluster is frequently produced in experiments despite its energetic metastability.

For the description of structural evolution of gold nanoclusters during the freezing and melting, we first introduce a structural order parameter $ \Phi ({\mathbf r}^N) $ that represents the degree of crystallinity in the system \cite{Steinhardt}. The free energy $F$ of the system with a particular value of structural order parameter $ \Phi $ at temperature $ T $ can be calculated by $ F(\Phi) = const - k_{\rm B} T \ln { P ( \Phi) } $, where $k_{\rm B}$ is the Boltzmann constant and $ P (\Phi) d \Phi $ is the probability to find the system with values of $\Phi$ between $\Phi$ and $\Phi+d \Phi$ \cite{ UnderstandingMD, Frenkel}.

In the above free energy calculations, conventional simulations \cite{Ohno} such as molecular dynamics (MD) simulations can be used in order to determine the probability $ P (\Phi) $. In this case, the MD events of the structural transformations rarely occur at the transition temperature for nanoclusters of over a few hundred atoms. As a result, the probability $ P (\Phi) $ is strongly peaked around a particular value of $\Phi$ corresponding to a certain equilibrium state at temperature $T$, while the probability is very low for a high free energy state. So it is practically impossible to obtain the free energy as a function of $ \Phi $ only with the conventional MD simulations. This problem, however, can be overcome by using the non-Boltzmann sampling such as the umbrella sampling. In the method, a $\Phi$-dependent bias potential makes the sampling probability appreciable in $\Phi$ region of high free energy \cite{UnderstandingMD, Frenkel, Wales}.

There are two factors that play a crucial role in the above biased sampling method: The first is a \emph{structural order parameter} $\Phi$ as a reaction coordinate that connects the initial and final states of the system, and the second is a $\Phi$-dependent \emph{bias potential} that forces the system to sample a high free energy region of $\Phi$ space along the reaction path.

\begin{table}
\caption
{\label{tab:table1}
Bond-odrer parameters $Q_6$ for various cluster structures.
}
\begin{ruledtabular}
\begin{tabular}{lcccc}
& liquid & Ih & Dh & fcc TOh \\
\hline
$Q_6$ & 0.06 & 0.17 & 0.30 & 0.57\\
\end{tabular}
\end{ruledtabular}
\end{table}
For the structural order parameter, we used the bond-orientational parameter $Q_6$, introduced by Steinhardt \emph{et al} \cite{Steinhardt}. This parameter, sensitive to the degree of orientational correlation defined by the vectors connecting neighbor atoms, measures the degree of crystallinity of the system. As shown in Table~\ref{tab:table1}, in a liquid cluster, $Q_6$ is very small (but nonzero due to the surface shape effect), while in a crystalline cluster, where the bond orientations are correlated coherently throughout the whole region, it is relatively large. We here note that $Q_6$ is not a state function with any specific internal structure but just a reaction coordinate indicating the degree of crystallinity.

For the fictitious bias potential of the umbrella sampling, we chose a harmonic function localized around a specific value of the order parameter $ \Phi (=Q_6 )$. We biased the configuration space sampling by adding this bias potential to the original potential energy of our system. By adjusting the center position of the harmonic function to the region of a large free energy barrier \cite{Frenkel, NHS1}, we could achieve relatively uniform sampling over the whole region of the phase transition. As a result, the transition between cluster phases of low and high crystallinity was made possible despite the substantial free energy barrier.

\begin{figure}
\includegraphics[width=0.37\textwidth]{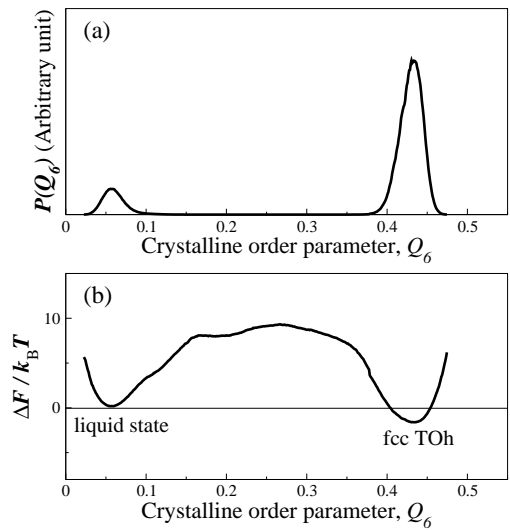} % Here is how to import EPS art
\caption
{ \label {fig:ProbabilityCurve}
Representative shapes of (a) probability distribution and (b) free energy ($\Delta F = F(Q_6 ) - F_{\rm liquid} $) as a function of a crystalline order parameter $Q_6$ at 780 K for a 459-atom Au
cluster. }
\end{figure}
We now investigate the behavior of freezing and melting of Au nanoclusters by calculating the free energies of the Au nanoclusters near the melting point. In these calculations, the probability $ P (\Phi) $ for the free energy was obtained by MD simulations based on the semi-empirical embedded-atom method \cite{EAM} and the bias potential scheme. As an initial cluster configuration, a TOh cluster of 459 atoms was used. The free energy was calculated at various temperatures between 760 and 820 K with an interval of 10 or 20 K by using the Anderson thermostat \cite{AndersonThermostat}. The space of bond-order parameter $Q_6$ was divided into 20 windows and umbrella sampling runs were performed with a bias potential corresponding to each window. Then the self-consistent histogram method \cite {UnderstandingMD} was used to reconstruct the overall distribution of probability $P(Q_6)$ from the individual histogram obtained in different windows. The total simulation time was more than $4 \times 10^5 \Delta t$, where $\Delta t=$2.5 fs represents the time step for the integration of the equation of motion. To check the reversibility of the free energy curves, we calculated the free energies by varying the $Q_6$ parameter in both increasing and decreasing directions. The calculations showed no significant hysteresis except at low temperature where the free energy changes steeply with $Q_6$. This indicates that our simulation time is long enough to equilibrate the system with negligible hysteresis. Typical results of the probability and free energy calculations are shown in Figs.~\ref{fig:ProbabilityCurve}(a) and (b), respectively. More details on the simulation procedures and numerical techniques are described elsewhere \cite{NHS1, InPreparation}.

\begin{figure}
\includegraphics[width=0.38\textwidth]{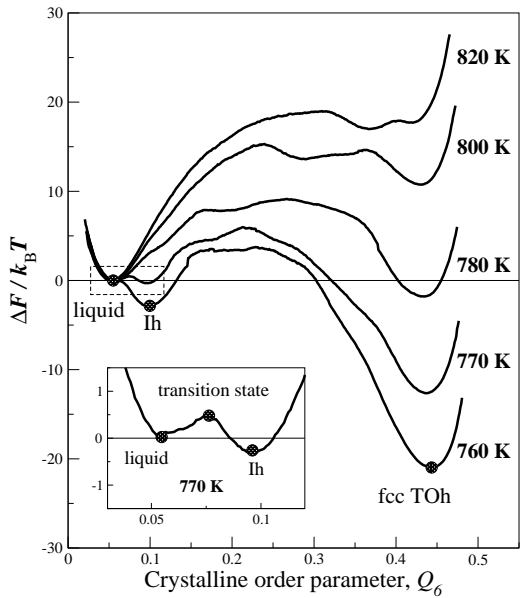} % Here is how to import EPS art
\caption
{ \label {fig:FreeEnergyCurve}
Free energy curves for a 459-atom Au cluster as a function of crystallinity ($Q_6$) at various temperatures. The free energy was obtained by using the umbrella sampling method combined with atomistic simulations. The inset shows a close-up of the curve at 770 K indicated by the box.}
\end{figure}

\begin{figure}
\includegraphics[width=0.160\textwidth]{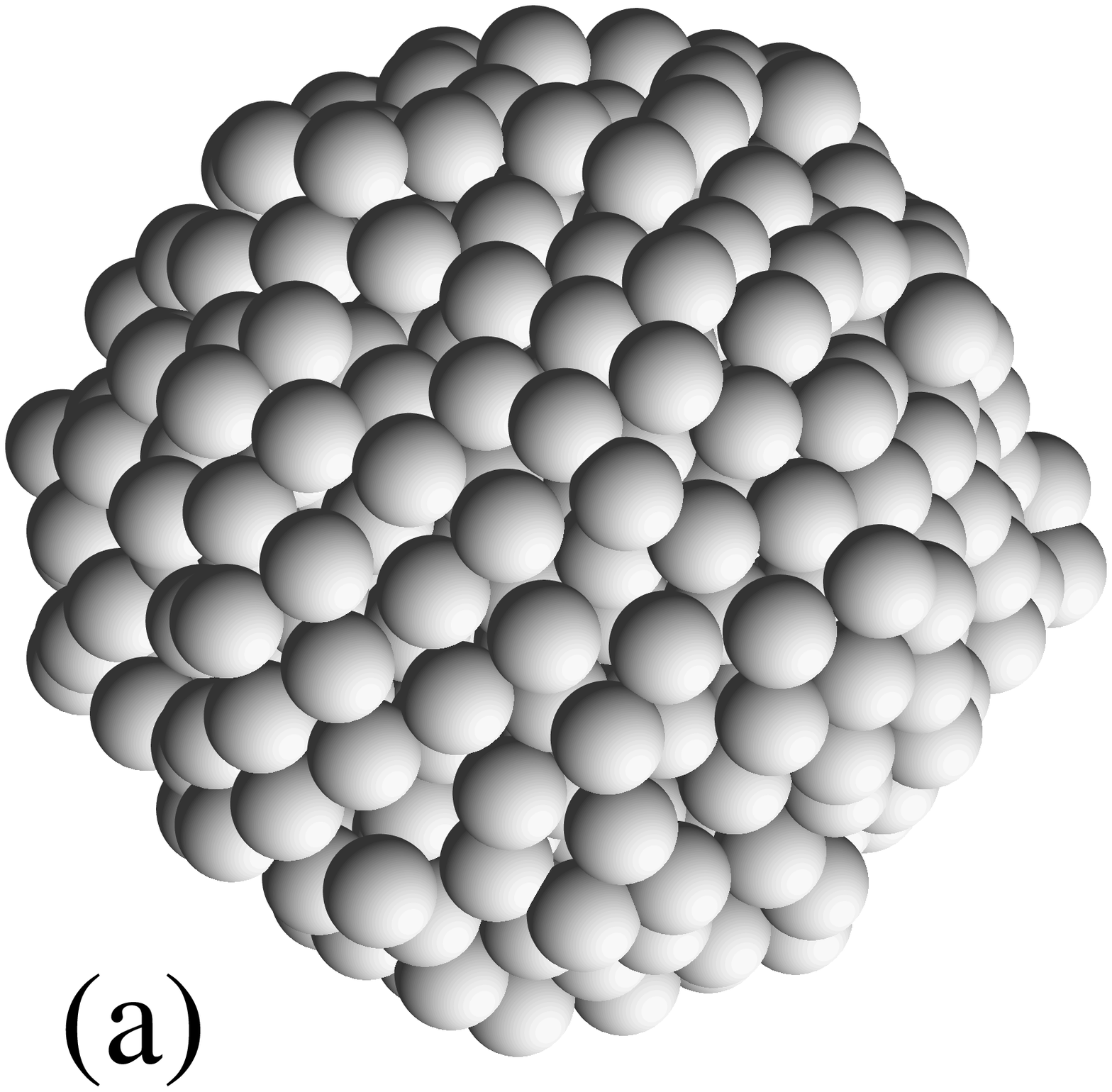}% Here is how to import EPS art
\includegraphics[width=0.158\textwidth]{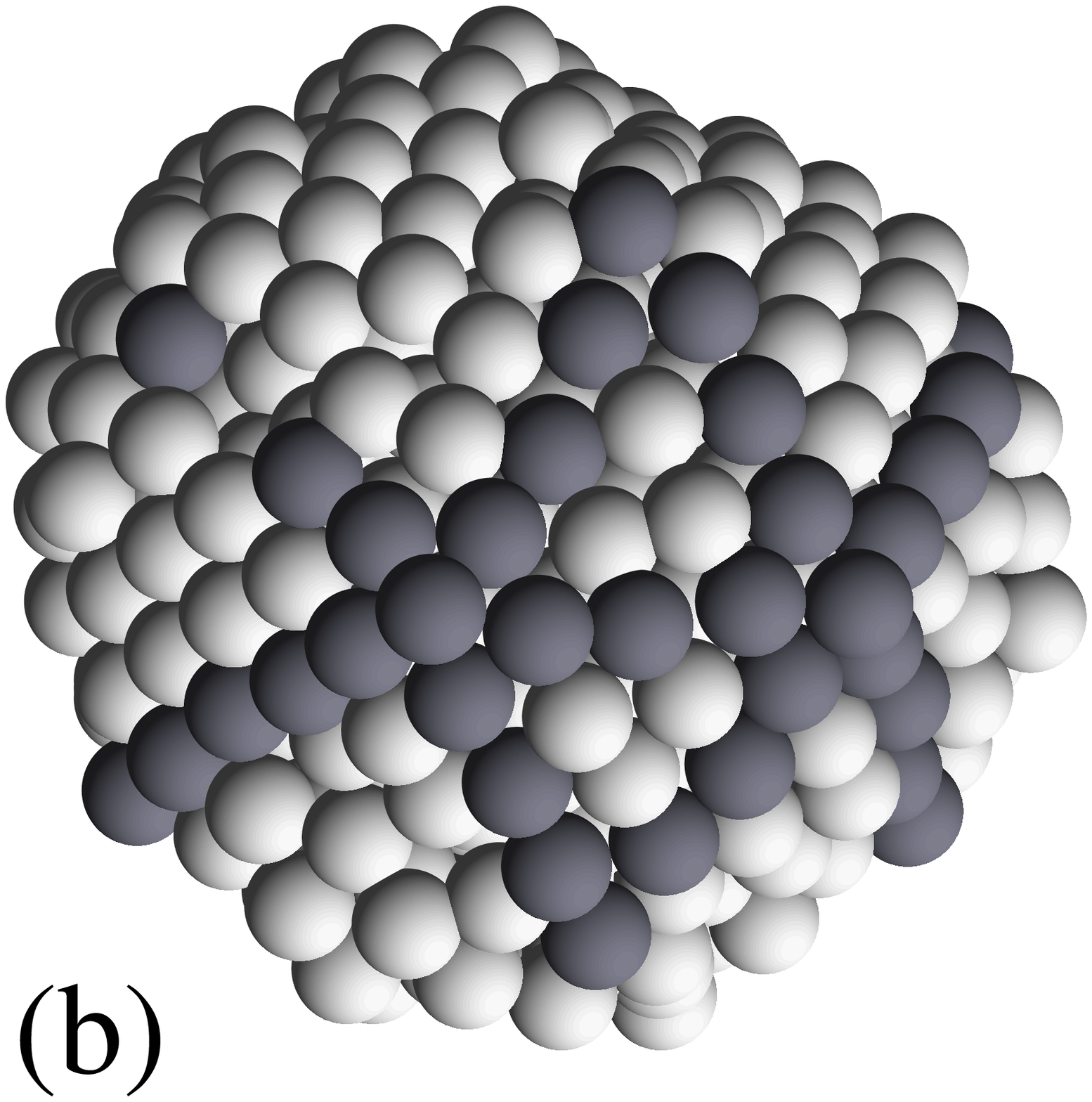}% Here is how to import EPS art
\includegraphics[width=0.162\textwidth]{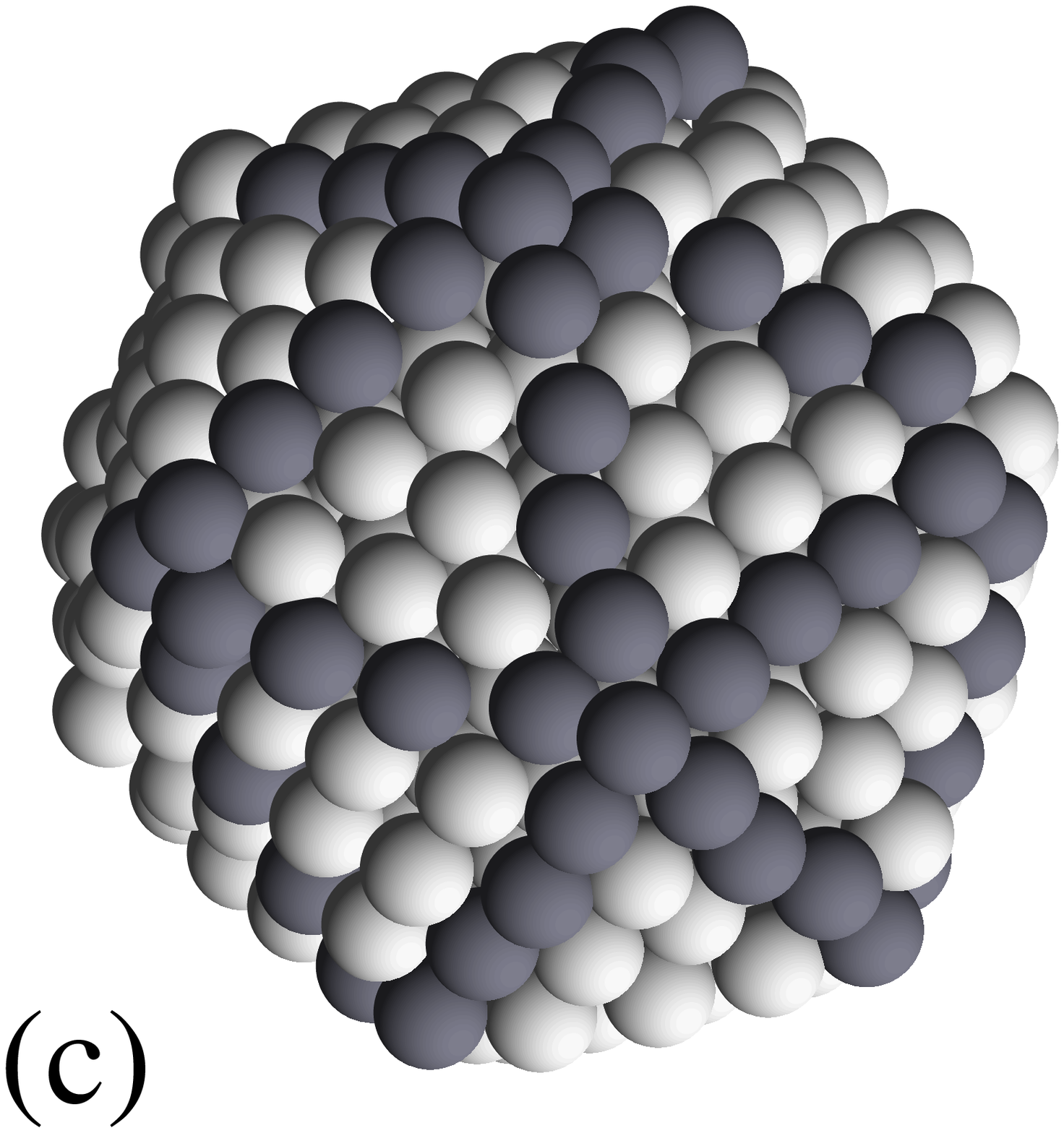}% Here is how to import EPS art

\includegraphics[width=0.16\textwidth]{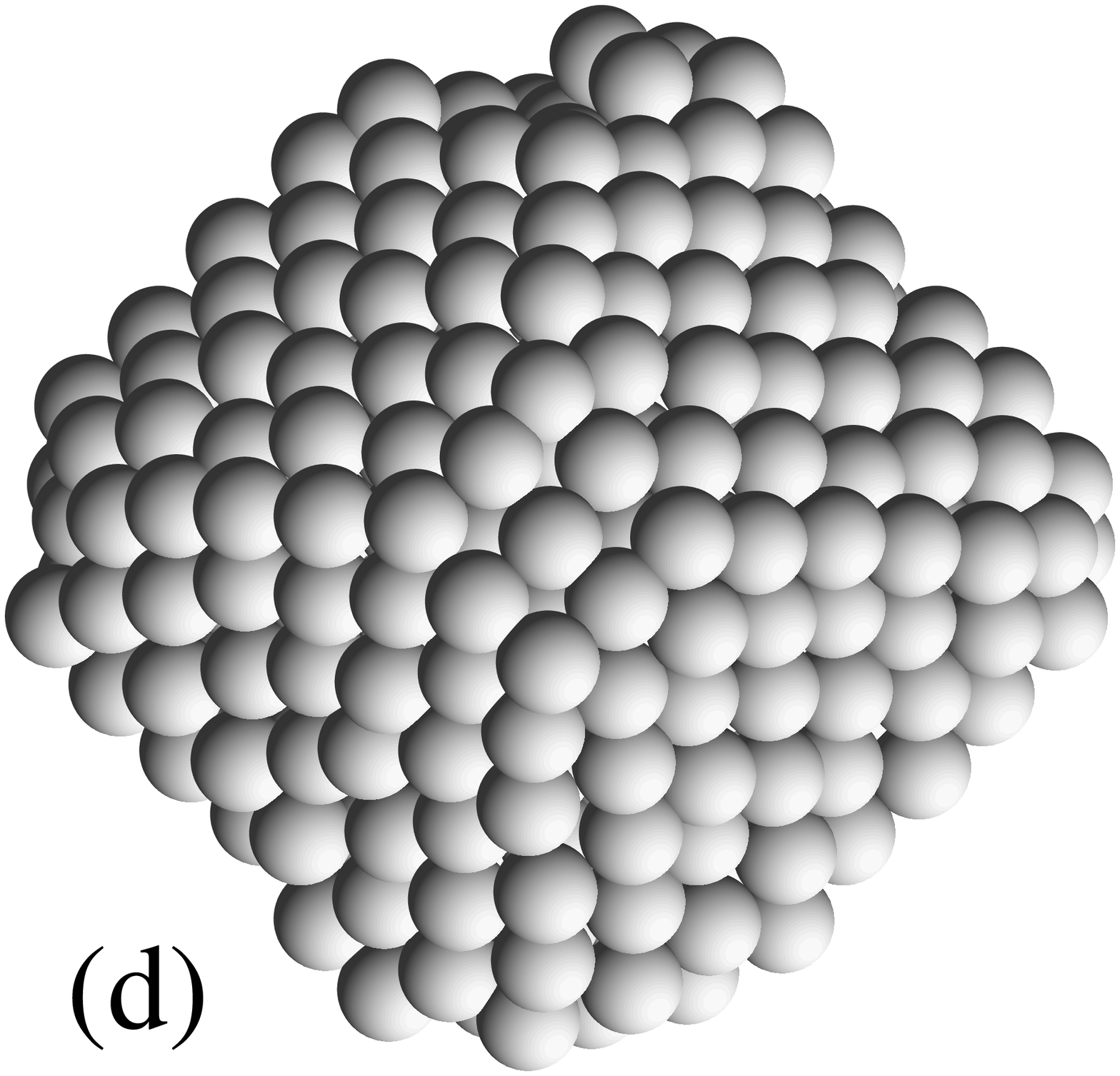}% Here is how to import EPS art
\hspace{0.3cm}%(e)
\includegraphics[width=0.29\textwidth]{fig3_inset}% Here is how to import EPS art
\caption { \label {fig:ClusterStructure} Representative atomic configurations of a 459-atom Au cluster at local free-energy-minima and transition state: (a) The liquid state around $Q_{6} \approx 0.05$, (b) the transition state between the liquid and the Ih state, (c) the Ih state around $Q_{6} \approx 0.1$, and (d) the TOh fcc crystalline state around $Q_{6} \approx 0.45$. (e) Calculated diffraction patterns for the structures of (a)-(c).}
\end{figure}

Figure~\ref{fig:FreeEnergyCurve} shows the free energy ($\Delta F = F(Q_6 ) - F_{\rm liquid} $) curves of the gold nanoclusters as a function of a crystalline order parameter $Q_6$. Here, the liquid state is taken as a reference free energy state. The free energy curves at various temperatures contain two or more energy minima representing the liquid and crystalline structures. In order to determine the cluster structures at the minimum states, the typical atomic configuration at each stage was captured as shown in Fig.~\ref {fig:ClusterStructure}. The two energy minima at $Q_6=0.45$ and $0.05$ represent the crystalline state of a TOh fcc structure and the liquid state, respectively. The phase transition temperature is around 780 K at which the free energies of solid and liquid are nearly equal. At high temperatures such as 820 or 800 K, the liquid state is a global energy-minimum, while the fcc solid state is metastable. Our free energy calculations show that at 820 K the initial fcc TOh cluster melts with a very low free energy barrier (less than a few $ k_{\rm B} T$). When temperature is further reduced, the relative free energy of the crystalline solid decreases monotonically and becomes a global minimum state at $T<780$ K.

Here, the free energy curves below the transition temperature are particularly interesting because they are expected to provide physical insight into the freezing behavior of the Au nanoclusters. Inspection of the curves gives two typical freezing paths, depending on the temperature. At small undercooling temperature of 780 K, the curve depicts a freezing path with a very large free energy barrier of 10 $k_{\rm B} T$ and shows the absence of any significant metastable solid state of a local free-energy minimum. However, at more undercooling temperature (770 K), the free energy changes significantly and a local free-energy-minimum state starts to appear at around $Q_{6} \approx 0.1$ [see the inset in Fig.~\ref{fig:FreeEnergyCurve}]. The typical atomic configuration at this state shows a fivefold symmetry on its surface [see Fig.~\ref {fig:ClusterStructure}(c)]. Further investigation of this metastable structure reveals that the cluster configuration corresponds precisely to the Ih structure only with some defects around the center of the cluster. At this temperature, the TOh fcc structure is thermodynamically the most stable. Here, it is noted that the free energy barriers are quite different: The free energy barrier for the transition from the liquid to the metastable Ih cluster is very low ($\sim k_{\rm B} T$), while the barrier from the liquid state to the fcc crystal is as large as 7 $k_{\rm B} T$. Interestingly, the typical atomic configuration at the transition state between the liquid and the Ih state shows \{$111$\}-type facets with a fivefold symmetry on its surface [see Fig.~\ref {fig:ClusterStructure}(b)], although the internal structure is rather liquid-like as shown in its diffraction pattern [see Fig.~\ref {fig:ClusterStructure}(e)]. As an atomic-level mechanism of structural transformation, we propose that the formation of the ordered facets provide a way of lowering the free energy barrier between the liquid and the Ih state \cite{NHS2}. This free energy behavior is more prominent with further decreasing of temperature (760 K) as shown in Fig.~\ref{fig:FreeEnergyCurve}. Considering the fast cooling rate of clusters in typical experimental conditions \cite{Bartell}, this simulation result clearly shows that the cluster can be dynamically trapped into the metastable state of the Ih structure with a very low free energy barrier during the freezing. The very low free energy barrier for the formation of the Ih structure explains why the Ih cluster structure is dominantly formed in spite of its energetic metastability \cite{AbsoluteRate}.

In addition to the freezing behavior, the melting of metal nanoclusters can be also explained by the free energy curves of Fig.~\ref{fig:FreeEnergyCurve}. Previous MD simulations reported that gold nanoclusters initially possessing an fcc single crystalline structure underwent a structural transformation to an Ih structure before melting \cite{ClevelandMD}. This result suggests that below the melting temperature, the Ih structure should be an intermediate stable state between an fcc single crystalline structure and the liquid phase \cite{ClevelandMD}. In contrast, our free energy curves in Fig.~\ref{fig:FreeEnergyCurve} show that the Ih structure is not stable along the melting path near the melting temperature. It is noted that the previous MD simulations were done under the constant energy condition \cite{ClevelandMD}. The latent heat effect of melting arising from the constant energy condition is expected to affect the structural transformation. Actually, the latent energy lowers the temperature by more than 80 K for this size of nanocluster. Such instantaneous refreezing of the cluster leads to the formation of the Ih local energy-minimum state before melting (see the free energy curve at 760 K of Fig.~\ref{fig:FreeEnergyCurve}). In order to confirm this argument, we compared MD simulations under constant energy and thermostat conditions. In accordance with the previous works, we obtained the formation of the Ih structure during melting under the constant energy condition, but we could not under the thermostat condition.

Wang \emph{et al.} \cite{Wang} also studied the melting behavior of Ih gold nanoclusters of a few thousand atoms and found that the cluster surface remains ordered up to the melting temperature, although it softens considerably with increased diffusion due to mobile vertex and edge atoms. This dynamics feature also agrees well with our observation that the high stability of \{$111$\}-type facets near the melting temperature lowers the free energy barrier for the transition between the liquid and the Ih cluster.

In our simulations, different cluster sizes can affect the freezing and melting behavior of nanoclusters. For example, we examined the cluster size effect by using a 561-atom cluster. For this cluster, the formation of the Ih structure is also dominant during freezing, in accordance with previous works \cite{Wang, Chushak, NHS2}. Further calculations of the free energies of the cluster clearly showed that the low free energy barrier from the liquid to the Ih phase contributes to the dominant formation of the Ih structure during freezing.

In summary, the freezing and melting behavior of gold nanoclusters was revisited in terms of the free energy as a function of a crystalline order parameter. By using the umbrella sampling technique combined with molecular dynamics simulations, the free energy barrier for the structural transition was calculated. It was found that the dominant formation of the Ih structure during freezing of Au nanoclusters is attributed to its low free energy barrier. Our free energy approach not only explains well the experimentally observed kinetics but also provides complementary information for understanding dynamical properties that are experimentally inaccessible, such as structural properties and formation dynamics of nano-size clusters. It opens a way to atomic simulations, capable of quantitatively studying the controlled growth of low dimensional nanostructures.

%%%%%%%%%%%%%%%%%%%%%%%%%%%%%%% Body of Paper %%%%%%%%%%%%%%%%%%%%%%%%%%%%%%%%%%

%-------------------- Acknowledgments ---------------------%

\begin{acknowledgments}
We gratefully acknowledge support from
the Korea Ministry of Science and Technology
through the Creative Research Initiative Program (H.-S.N. and D.-Y.K.),
the Korea Ministry of Education
through the Brain Korea $21$ Program (N.M.H. and J.-K.Y.),
and the Korea Research Foundation (KRF)
under Agreement No. KRF-2003-015-C00210 (B.D.Y.).
Fruitful discussions with J.W. Lee and G-D. Lee at Seoul National University are appreciated.
\end{acknowledgments}

%\newpage %Just because of unusual number of tables stacked at end

%%%%%%%%%%%%%%%%%%%%%%%%%%%%%%% BIBLIOGRAPHY %%%%%%%%%%%%%%%%%%%%%%%%%%%%%%%%%%%

% \bibliography{Ih_PRL}% Produces the bibliography via BibTeX.

\begin{thebibliography}{99}
\bibitem{AbsoluteRate}
The absolute rate of transtion $J$ can be described by
$J = A \exp (-\Delta F / k_{\rm B} T )$,
where A is a kinetic prefactor whose optimum form is not entirely resolved.
However, the exponential term contributes dominantly to the relative rate of transition.
\end{thebibliography}

\end{document}